# Electronic and optical properties of SnX$_2$ (X=S, Se) - InSe van der Waal's heterostructures from first-principle calculations

Amretashis Sengupta[1,2*†]

[1]Independent Researcher, 5 Bagha Jatin Park, Road-II, Siliguri, WB-734001, India.

[2]Indian Institute of Engineering Science and Technology, Shibpur, Howrah – 711 103, India

**Corresponding author, E-mail:** amretashis@gmail.com

***Abstract:*** *In this work from first-principles simulations we investigate bilayer van der Waal's heterostructures (vdWh) of emerging 2-dimensional (2D) optical materials SnS$_2$ and SnSe$_2$ with monolayer InSe. With density functional theory (DFT) calculations, we study the structural, electronic, optical and carrier transport properties of the SnX$_2$ (X=S,Se)-InSe vdWh. Calculations show SnX$_2$-InSe in its most stable stacking form (named AB-1) to be a material with a small (0.6-0.7eV) indirect band-gap. The bilayer vdWh shows broad spectrum optical response, with number of peaks in the infra-red to visible region. In terms of carrier transport properties, asymmetry in conductance was observed with respect to the transport direction and electron and hole transmission. The findings are promising from the viewpoint of nanoelectronics and photonics.*

## I. Introduction

Layered tin dichalcogenides such as SnS$_2$ and SnSe$_2$ have recently been in focus for their novel properties [1]-[6] such as indirect to direct band-gap transition [1,5], strain tunable magnetism [3], high carrier mobility [2], low thermal conductivity [2], fast photoresponse [1] etc. These properties indicate wide-ranging potential applications of 2-dimensional (2D) SnX$_2$ (X=S,Se) in in sensors, nanoscale FETs, thermoelectrics, catalysis, solar cells and photonic devices. [4,5,6] The industrial viability of these materials are enhanced by the choice of large number of chemical and physical synthesis techniques for layered SnX$_2$ [1]-[6] and the relative abundance and environmental friendliness of the constituent elements [1].

Also, another emerging material, layered InSe has shown good band-gap tunability and carrier transport properties [7]-[12] and good photo-detection and novel optical response wherein the band edge response is absent in monolayers [7,8]. As a result layered or 2D InSe is also being pursued for nanoelectronics and photonics in a significant manner. [11,12]

---

[†]The author was with the Indian Institute of Engineering Science and Technology, Shibpur at the time of carrying out this work. Presently he is an independent researcher not affiliated to any institution.

Possessing excellent lattice matching, 2D hexagonal $SnX_2$ and InSe have very good potential compatibility in sequentially grown superlattices or van der Waals heterostructures (vdWh). Such a combination, could offer interesting optical and electronic properties, owing to the close values of band-gaps resulting in a high possibility of intra and inter-band transitions among the two different materials.

In this work we look to investigate $SnX_2$-InSe vdWh from first principle density functional theory (DFT) simulations. We identify the most stable stacking of the vdWh lattice and thereafter calculate the electronic properties and various optical properties as joint density of state (JDOS), complex dielectric functions and energy loss spectra (theoretical electron energy loss spectra or EELS), of the same. Also we investigate the carrier transport properties in such vdWh by means of non-equilibrium Green's function (NEGF) simulations.

## II. Methodology

We employ generalized gradient approximation (GGA) calculations with Perdew-Burke-Ernzerhoff (PBE) exchange and correlation functional [13] in the Quantum ESPRESSO package [14]. The vdWh unit cells consisting of $SnX_2$-InSe had 15Å vacuum gap, and they were sampled with a 9x9x1 Monkhorst-Pack [15] k-point grid, with a cutoff energy of 150Ry. The Davidson diagonalization algorithm [16] was used in the calculations, with convergence threshold of $10^{-7}$ Ry. The fhi88pp norm-conserving basis sets of Troullier-Martins type [17,18] were used in the calculations. The van der Waals corrections were included by the Grimme's DFT-D2 method [19] and the structures were relaxed with Broyden-Fletcher-Goldfarb-Shanno (BFGS) algorithm [20] for variable cell relaxation to a pressure convergence threshold of 0.5kBar and with Hellmann-Feynman forces reduced to less than $10^{-4}$Ry/Bohr.

The optical spectra of the systems were calculated with the random phase approximation (RPA) approach, as implemented in the epsilon.x toolset [21] The joint density of states (JDOS) is defined as [21]

$$J(\omega) = \sum_{\alpha} \sum_{\beta} \frac{V}{(2\pi)^3} \int \delta\left(E_{k,\beta} - E_{k,\alpha} - \hbar\omega\right) \left[ f(E_{k,\alpha}) - f(E_{k,\beta}) \right] \cdot d^3k \qquad (1)$$

$V$ is the volume of the cell, $\alpha, \beta$ are states belonging to conduction and valence bands respectively, $E_{k,}$ are the eigenvalues of the Hamiltonian and $f(.)$ is the Fermi distribution function. The Dirac delta function is approximated by a Gaussian distribution, normalized to one [21]

$$G(\omega) = \frac{1}{\Gamma\sqrt{\pi}} \exp\left\{\left(E_{k,\beta} - E_{k,\alpha} - \hbar\omega^2\right)/\Gamma^2\right\} \qquad (2)$$

$\Gamma$ is the broadening parameter. The complex dielectric function is given by [21]

$$\varepsilon_2(\omega) = 1 - \frac{4\pi q^2}{VN_k m^2} \sum_{\alpha,k} \frac{df(E_{k,\alpha})}{dE_{k,\alpha}} \frac{\hat{M}_{a,b}}{\omega^2 + i\eta_1\omega} + ...$$
$$... + \frac{8\pi q^2}{VN_k m^2} \sum_{\alpha \neq \beta} \sum_k \frac{\hat{M}_{a,b}}{(E_{k,\beta} - E_{k,\alpha})} \frac{f(E_{k,\alpha})}{(\omega_{k,\beta} - \omega_{k,\alpha})^2 + \omega^2 + i\eta_2\omega}$$

(3)

$\eta_1$ and $\eta_2$ are the inter-band and intra-band smearing respectively. The matrix element $M_{a,b}$ are defined as [21]

$$\hat{M}_{a,b} = \left\langle u_\beta \mid p_a \mid u_\alpha \right\rangle \left\langle u_\alpha \mid p_b^\dagger \mid u_\beta \right\rangle$$

(4)

With a London transformation upon $\varepsilon_2$ the dielectric tensor calculated on the imaginary frequency axis may be written as [21]

$$\varepsilon_2(i\omega) = 1 + \frac{2}{\pi} \int_0^\infty \frac{\omega' \varepsilon_2(\omega')}{\omega'^2 + \omega^2} d\omega'$$

(5)

The energy loss spectrum (theoretical EELS count) is obtained from the imaginary part of the inverse dielectric tensor.

The transmission is calculated with the DFT-NEGF method [22]-[28] in QuantumWise ATK 2016.4 [26], using similar DFT simulation parameters as with the Quantum ESPRESSO calculations. For the NEGF transport calculation, an FFT2D Poisson solver and Krylov self-energy calculator are employed in ATK. [26]-[28] The detailed DFT-NEGF transport calculation methodology is well-described in literature [22]-[28] The zero-bias transmission spectra were evaluated for both the zigzag and the armchair direction. The Green's function can be expressed as [22]-[28]

$$G(\mathrm{E}) = \left[ (\mathrm{E} + i\delta)S - H - \Sigma^R(\mathrm{E}) - \Sigma^L(\mathrm{E}) \right]^{-1}$$

(6)

In (X) $H$ is the Hamiltonian, and $S$ is the overlap matrix. $\delta$ is an infinitesimally small positive quantity. The self-energy matrices of the right/left contacts is $\Sigma^{R,L}$ and is related to the broadening matrix $\Gamma$ as [22]-[24]

$$\Gamma = i\left(\Sigma - \Sigma^\dagger\right)$$

(7)

The transmission can be calculated from the Green's function as [22,23,28]

$$T(\mathcal{E}) = \mathrm{Tr}\left[ G\Gamma^R G^\dagger \Gamma^L \right]$$

(8)

## III. Results and Discussions

The hexagonal monolayer SnX$_2$ has the Sn atom sitting at the centre of edge sharing octahedral and lying between two layers of chalcogen atoms. Such a structure when combined with the corrugated surface of the InSe monolayer, presents various possible configurations. Considering the three high symmetry points in the hexagonal lattice and three dissimilar atoms in the SnX$_2$ layer (in terms of element as well as position), and the corrugated nature of the InSe with two (chemically) distinguishable crystallographic positions, there can be in total 12 different possible structures of the SnX$_2$-InSe vdWh. For ease of understanding the difference between subsets of the three different stackings, namely AA, AB and AC, we use a labelling scheme follows. The top chalcogen atom, bottom chalcogen atom and the metal atom in SnX$_2$ as c1t, c1b and m1 respectively, whereas the top In atom and top Se atoms are labelled m2 and c2 respectively. From the point-of-view of stacking, as the bottom In and Se atoms share the same in-plane co-ordinates, they need not be labelled differently to distinguish variations in stacking. In the current naming scheme AA-c1bc2-c1tm2, implies AA stacking with the c1b atom lying on top of c2 and c1t on top of m2.

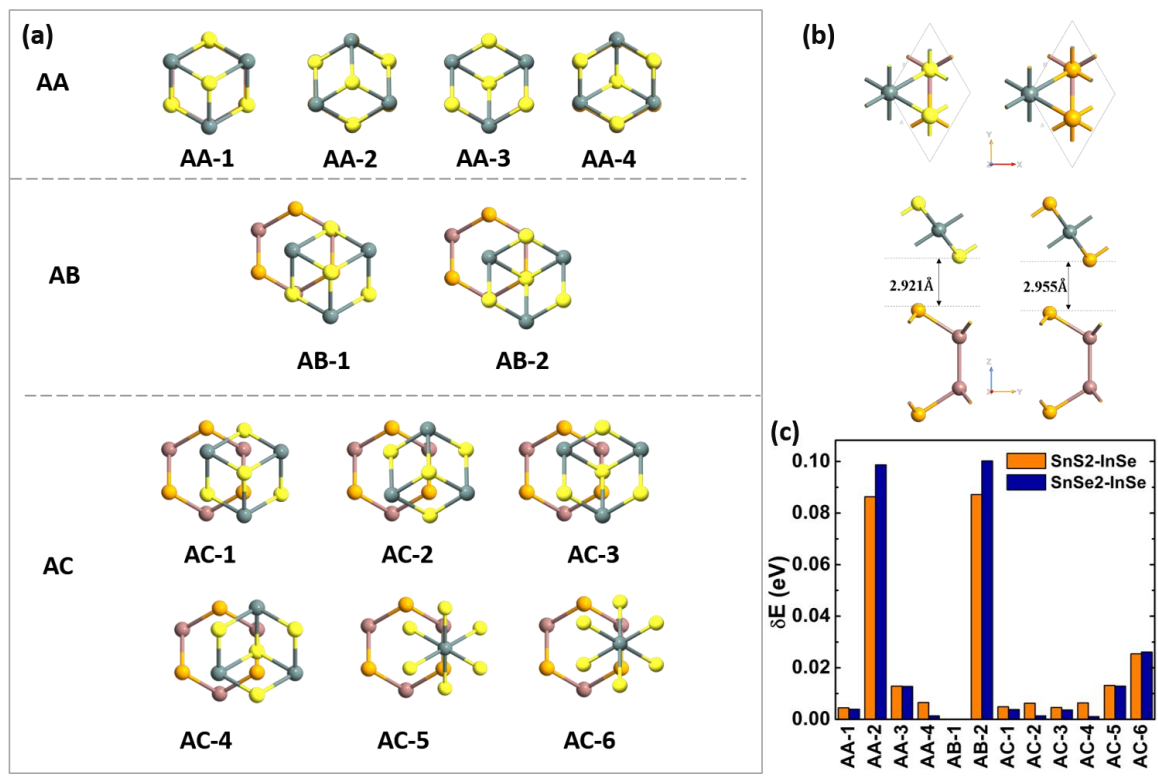

**Fig. 1: (a)** The top views (xy-plane) of the different possible stacking structures structures of the SnX$_2$-InSe vdWh **(b)** The top (*xy*-plane) and side (*yz*-plane) view of the most stable structure (AB-1) and **(c)** the relative total energies with respect to the most stable structure.

The optimized cell parameters of the different structures are provided in Table I. Of the different structures the AB-c1tc2-c1bm2 (which we will refer to in short as AB-1), is the most stable, for both SnS$_2$-InSe and SnSe$_2$-InSe vdWh structures. While there are other structures as well which differ only by a very small amount of energy in terms of stability, we look to focus mostly on the most energetically stable configuration in the further studies.

**Table I:** Different structures of the SnX$_2$-InSe vdWh and associated lattice parameters and interlayer spacing.

| Stacking | Short Name | Lattice parameter a (Å) | | Interlayer spacing d (Å) | |
|---|---|---|---|---|---|
| | | SnS$_2$-InSe | SnSe$_2$-InSe | SnS$_2$-InSe | SnSe$_2$-InSe |
| AA-c1bc2-m1m2 | AA-1 | 3.8139 | 3.8894 | 2.9777 | 2.9704 |
| AA-c1bm2-m1c2 | AA-2 | 3.8098 | 3.8802 | 3.7131 | 3.7756 |
| AA-c1tc2-m1m2 | AA-3 | 3.8139 | 3.8919 | 2.9777 | 2.9772 |
| AA-c1tm2-m1c2 | AA-4 | 3.8138 | 3.8868 | 2.9758 | 2.9375 |
| AB-c1tc2-c1bm2 | AB-1 | 3.8133 | 3.8912 | 2.9213 | 2.9548 |
| AB-c1tm2-c1bc2 | AB-2 | 3.8086 | 3.8801 | 3.7196 | 3.7849 |
| AC-c1tx-u | AC-1 | 3.8114 | 3.8846 | 3.0642 | 3.1113 |
| AC-c1bx-d | AC-2 | 3.8115 | 3.8856 | 3.037 | 3.0857 |
| AC-c1tx-u | AC-3 | 3.8113 | 3.8853 | 3.0681 | 3.1131 |
| AC-c1bx-d | AC-4 | 3.8119 | 3.8858 | 2.8639 | 2.9106 |
| AC-m1x-u | AC-5 | 3.8154 | 3.8843 | 3.0677 | 3.1124 |
| AC-m1x-d | AC-6 | 3.8056 | 3.8766 | 3.091 | 3.1488 |

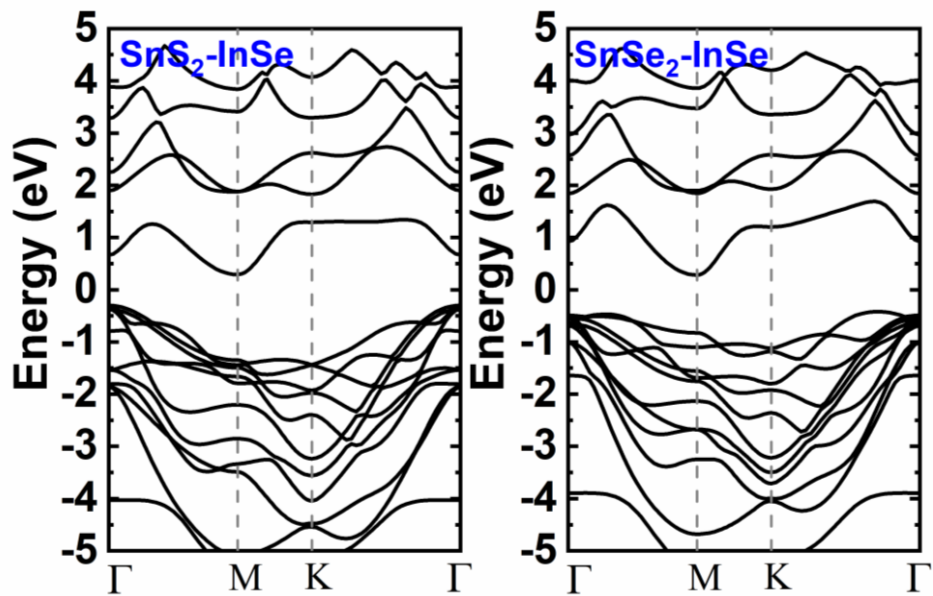

**2. (a)**

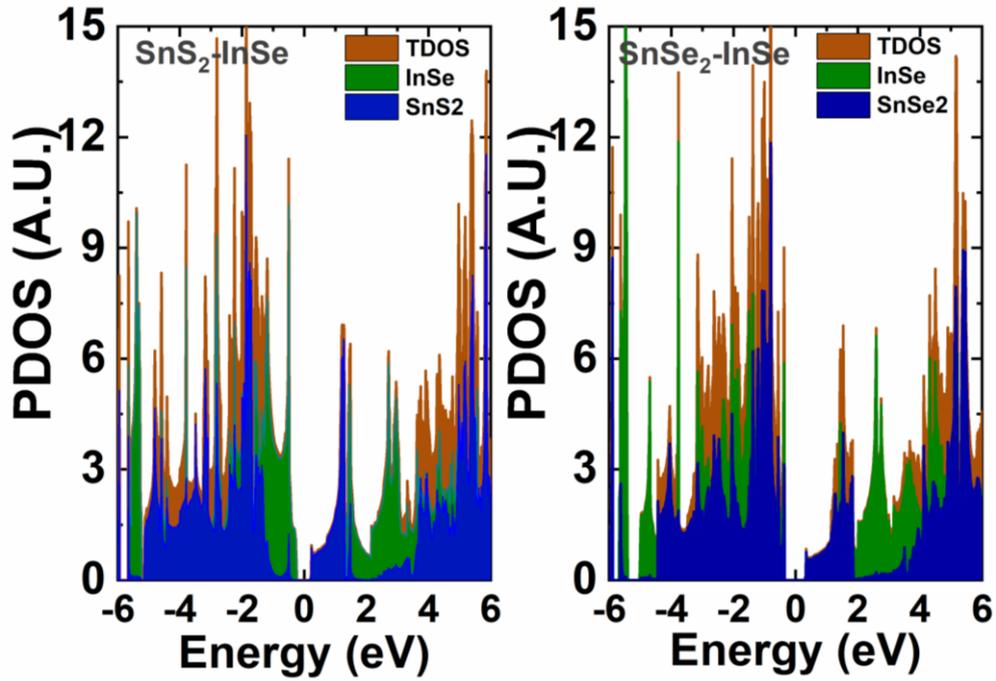

**2. (b)**

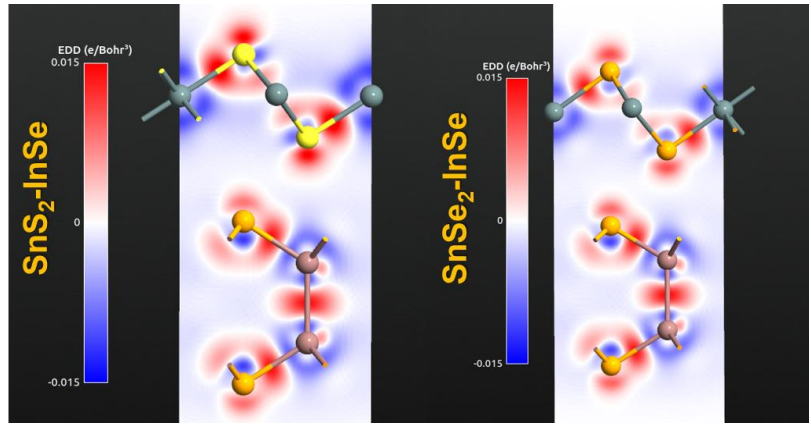

**2. (c)**

**Fig. 2 : (a)** PBE Bandstructures of the SnX$_2$-InSe vdWh, showing also the contribution of the comprising layers in dashed lines. **(b)** Projected Density of states (PDOS) plots showing the contributions of the individual layers and the resulting total DOS in the vdWh. **(c)** Electron density difference (EDD) plots for the vdWh

In Fig. 2a, the bandstructures of the most stable form (AB-1 stacking) of the SnS$_2$-InSe and the SnSe$_2$-InSe vdWh are presented. Both the vdWh show indirect band-gap. SnS$_2$-InSe shows a gap of 0.61eV between M point in conduction band (CB) and Γ point in valence band (VB). SnSe$_2$-InSe has a slightly larger gap of 0.70, with the CB minima at M point and VB maxima at a point in the K-Γ path, located about two third the distance from K to Γ point.

From the bands in Fig. 2a, we see that the conduction band (CB) minima in both the vdWh are located at the M point of the hexagonal Brillouin zone. The PDOS plots in Fig. 2b show the vdWh conduction bands to be more influenced by the $SnX_2$ layer. However for the valence band (VB) maxima, the $SnX_2$ contribution seems to depend on chalcogen of the top layer (X=S or Se). PDOS reveals that it is InSe that has more contribution to VB edges in $SnS_2$-InSe as compared to $SnS_2$, while for $SnSe_2$-InSe vdWh, it is $SnSe_2$ which contributes more significantly to the VB states. These different type of contributions push the VB max upwards at $\Gamma$ point in case of $SnS_2$-InSe and somewhere between the K and $\Gamma$ points for $SnSe_2$-InSe vdWh.

The difference in the charge distribution or transfer in the two vdWh is investigated with electron density difference (EDD) analysis, presented in Fig 2c. The more positive EDD around the chalcogen atoms of the $SnX_2$ in case of $SnS_2$-InSe, as compared to $SnSe_2$-InSe, indicates a stronger tendency for electron accumulation near the S atoms in $SnS_2$ than the Se atoms in $SnSe_2$ vdWh. Apart from this a tendency of developing a stronger dipole moment between the Sn-S pair as compared to the Sn-Se pair is also evident from the EDD plot. These factors revealed by the PDOS and the EDD results contribute to the bandstructure differences among the two vdWh.

The carrier effective masses ($m^*$) for the vdWh were calculated with a polynomial fitting formula, considering the parabolic band approximation and using the relation $\frac{1}{m^*} = \frac{1}{\hbar^2}\left(\frac{d^2\xi}{dk_i dk_j}\right)$ , $\xi$ being the energy eigenvalue and $k_{i,j}$ the wave-vector along $i, j$ direction at the VB maxima and the CB minima in the (100) and (010) directions. The calculated effective masses are listed in Table-II.

**Table II:** The calculated band gap and carrier effective mass for the vdWh

| vdWh | Band gap $E_g$ (in eV) | electron effective mass ($m_e^*$) | | hole effective mass ($m_h^*$) | |
|---|---|---|---|---|---|
| | | (100) | (010) | (100) | (010) |
| $SnS_2$-InSe | 0.61 (Indirect) | 0.572 $m_0$ | 0.675$m_0$ | 0.897 $m_0$ | 0.795$m_0$ |
| $SnSe_2$-InSe | 0.70 (Indirect) | 0.653 $m_0$ | 0.789$m_0$ | 0.956 $m_0$ | 0.992$m_0$ |

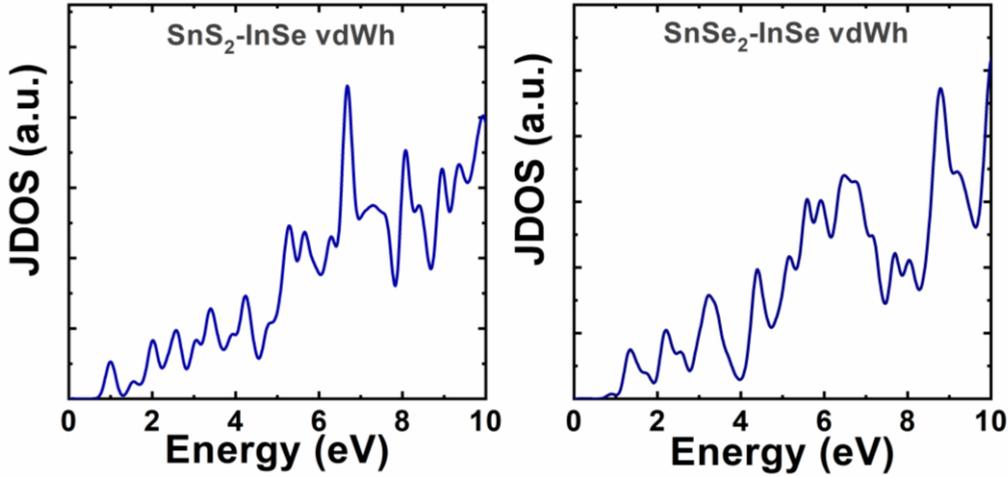

**Fig 3:** The calculated joint density of states (JDOS) of the SnX$_2$-InSe vdWh.

The calculated optical joint density of states (JDOS) of the vdWh are presented in Fig. 3. As JDOS is an indicator of the number of available states for photons to interact with, for optical emission/absorption process, it is an important part of optical characteristics of a given material. The JDOS of the two structures differ in the sharpness of the peaks apart from their position and also the number of peaks. While the SnS$_2$-InSe shows a greater number of sharper peaks, the JDOS peaks for SnSe$_2$-InSe are more spread-out in nature, and also lesser in number. A cascade of small but well defined peaks in the range 0.5-5eV for SnS$_2$-InSe vdWh outnumbers those in SnSe$_2$-InSe. Around the 1eV energy range, SnS$_2$-InSe shows more possible transitions than SnSe$_2$-InSe. Also around a higher energy of 6.5-7.5eV, there exists a significant difference in terms of the JDOS among the two vdWh. The SnSe$_2$-InSe vdWh is characterised by a significant decline in the JDOS around the 7-8eV energy range.

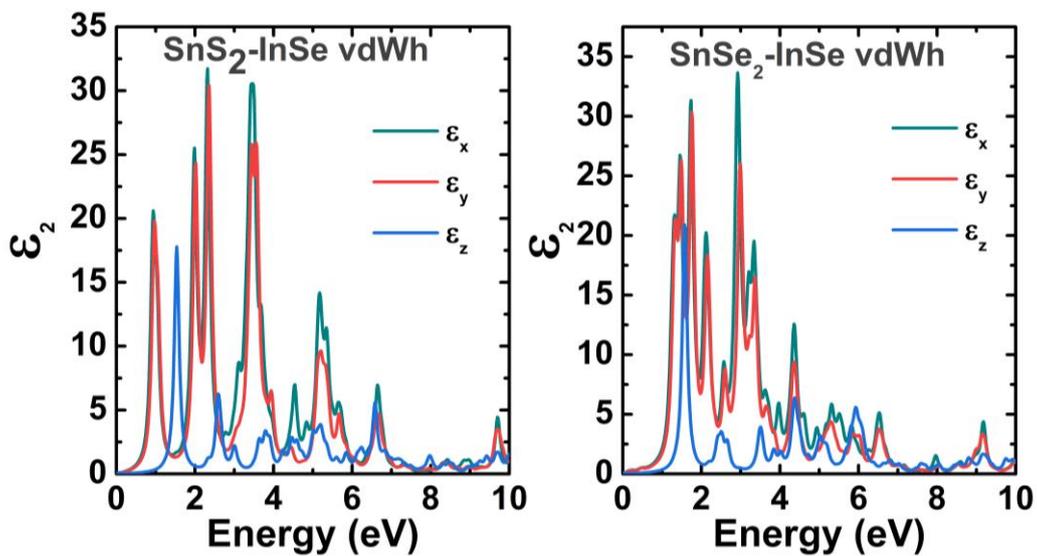

**Fig 4:** The calculated imaginary part of the complex dielectric function of the SnX$_2$-InSe vdWh.

The imaginary part of the complex dielectric function ($\varepsilon_2$), which can also act as a measure of optical absorption, is presented in Fig. 4. The two in-plane components (*x* and *y*) and the out-of plane (z) component are all shown in the plots. A stronger absorption in the infrared (IR) to visible range is observed for both the vdWh. For both the structures, the *z*-component of $\varepsilon_2$ appears blue-shifted as compared to the in-plane (*x* and *y* components). The blue-shift is more prominent in case of the SnS$_2$-InSe vdWh. Considering the position of the in-plane characteristics, more discrete peaks at lower energies between 0.75-2.5eV can be observed for SnS$_2$-InSe as compared to SnSe$_2$-InSe vdWh. The peaks in case of the SnSe$_2$-Inse vdWh are more concentrated together, in the lower energy ranges. These corroborate with the JDOS spectra discussed previously. The spike in available states in SnS$_2$-InSe for transition around the 6.5eV mark, is also well reflected by the presence of a stronger response in this region by the SnS$_2$-InSe vdWh, as compared to SnSe$_2$-InSe vdWh. Here it must be mentioned that the results presented herein are based on a random-phase approximation (RPA) method as described by Benassi et. al. [21]. This implementation based on a multi-particle approach and offers superior accuracy over simulations with the single particle dielectric tensor. However it does not include the excitonic effects for which GW and Bethe-Salpeter Equation (BSE) calculations could yield better results. However with the present level of implementation, the results presented herein can be considered a good pointer towards the optical response of the vdWh systems under consideration.

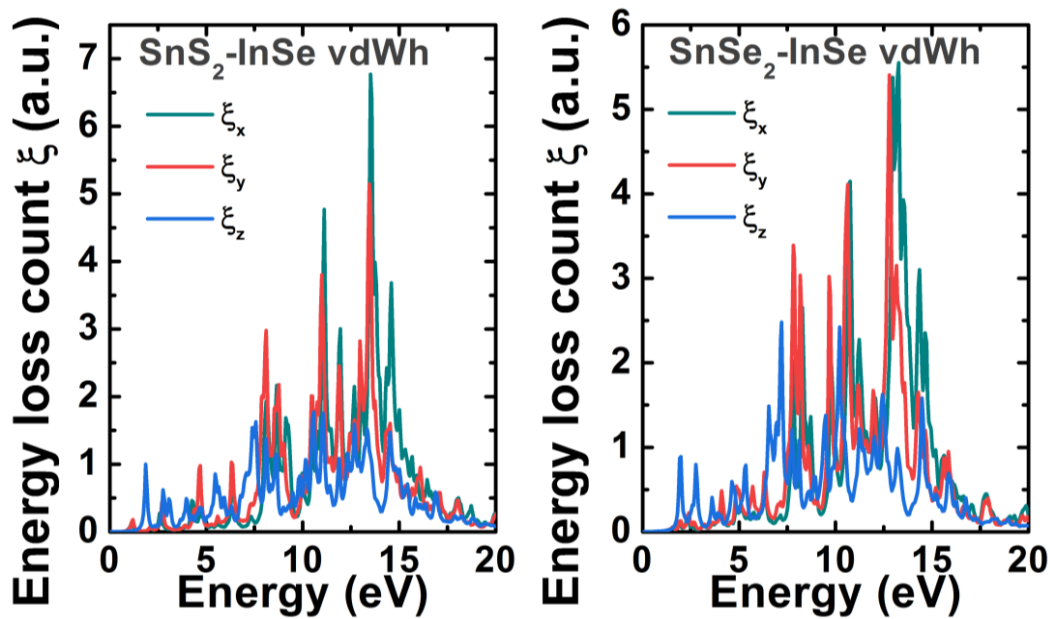

**Fig 5:** The energy loss spectra of the SnX$_2$-InSe vdWh.

The theoretical EELS count is presented in Fig. 5, shows the maxima for the SnS$_2$-InSe vdWh at 13.27, 12.79 and 7.21eV for the *x, y* and *z* components respectively. For the SnSe$_2$-InSe vdWh, the same are observed at 13.51, 13.45 and 10.57eV. The bulk plasmon frequency for these systems can thus be estimated to be $8.94 \times 10^4$ cm$^{-1}$ and $1.01 \times 10^5$ cm$^{-1}$ for SnS$_2$-InSe and SnSe$_2$-InSe vdWh respectively.

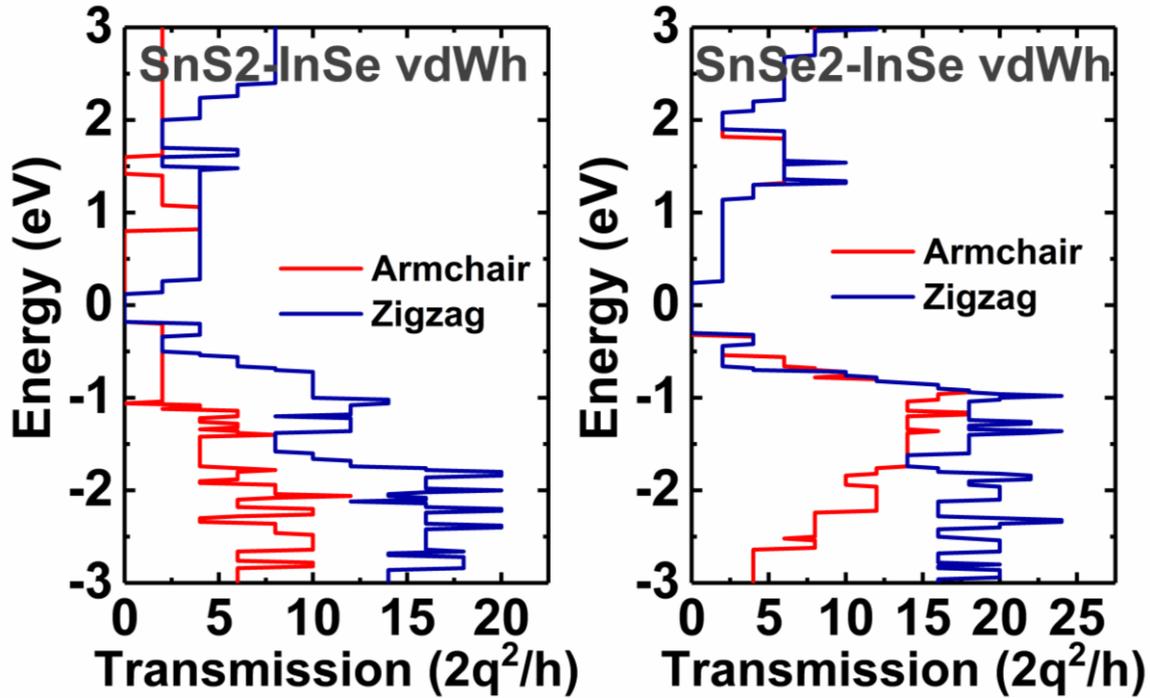

**Fig 6:** The carrier transmission spectra of the SnX$_2$-InSe vdWh.

The transmission spectra of both of the vdWh show an enhanced carrier transport properties for the valence bands rather than the conduction bands. Also observable is the asymmetry in conductance in the armchair and the zigzag direction of the SnX$_2$-InSe structures. The suppression of transport in the armchair direction than that in the zigzag direction is more significant for the SnS$_2$-InSe vdWh, than the SnSe$_2$-InSe vdWh. However the magnitude of transmission for holes in both transport directions is more in case of the SnSe$_2$-InSe structure. For electron transport there exist very little asymmetry on the transport direction for SnSe$_2$-InSe vdWh, as compared to SnS$_2$-InSe, additionally the magnitude for the zigzag direction, is also quite close for both the vdWh. While the asymmetry may be attributed to the disparity in carrier effective masses in the different transport directions, the enhanced hole conduction is a result of the larger number of valence states available near the Fermi-level (as visible in the DOS). The asymmetry in terms of transport direction and electron/ hole transmission, is something that can be utilized for novel applications in photonic and electronic devices based on SnX$_2$-InSe vdWh.

# IV. Conclusion

In this work by density functional theory simulations the structural, electronic and optical properties of bilayer van der Waals heterostructures made of 2D $SnX_2$ (X=S,Se) and InSe were investigated. Among the 12 different possible stacking combinations AB-1 stacking, was found to be the most stable for both the vdWh. The stable vdWh structures were thereafter studied for their bandstructure, DOS, JDOS and other optical parameters such as the imaginary dielectric function and theoretical EELS spectra. It was observed that $SnS_2$-InSe and $SnSe_2$-InSe vdWh form indirect-gap materials with small PBE band-gap values of 0.61 and 0.70eV respectively. The vdWh also showed moderate electron and hole effective masses ranging between $0.57-0.78m_0$ and $0.79-0.99m_0$ respectively depending upon the material and the transport directions. A good optical response could be observed in the IR-visible range for both the $SnX_2$-InSe vdWh, with the peaks in the dielectric function being more closely packed and also slightly blue shifted for the $SnSe_2$-InSe as compared to $SnS_2$-InSe. A drop in JDOS and optical response could be observed around 6-6.5eV and 6.75eV for the $SnSe_2$-InSe, as compared to its $SnS_2$ counterpart. The bulk plasmon frequency was estimated to be around $8.94x10^4$ and $1.0x10^5$ $cm^{-1}$ for $SnS_2$ and $SnSe_2$ based vdWh respectively. Carrier transport computed with the DFT-NEGF method, showed asymmetric conductance in terms of transport direction (armchair / zigzag) and the nature of carriers (electrons/ holes). The optical and transport properties calculated show good prospect of application of $SnX_2$-InSe vdWh studied in this work, in photonics and nanoelectronics. The excellent lattice matching and environmental stability of the constituent 2D materials, make a good case for not only bilayer but for multi-layer $SnX_2$-InSe devices.

## V. Acknowledgements

The author wishes to thank Department of Science and Technology, Govt. of India for DST INSPIRE Faculty Grant No. IFA-13 ENG- 62.